\documentclass[11pt,a4paper]{article} 

\usepackage[scaled]{helvet}
\usepackage[T1]{fontenc}

\usepackage{amsthm, amsmath, amssymb, color, bm}
\usepackage[authoryear]{natbib}
\usepackage[utf8]{inputenc}
\usepackage{enumitem}
\usepackage{graphicx}
\usepackage{authblk}
\usepackage[left=0.75in,right=0.75in,top=1in,bottom=1in]{geometry} 
\usepackage{setspace} 

\title{Decisions, decisions, decisions in an uncertain environment\vspace{-0.5em}}
\author[1]{Noel Cressie (ncressie@uow.edu.au)\vspace{-0.5em}}
\affil[1]{NIASRA, School of Mathematics and Applied Statistics, University of Wollongong, Australia\vspace{-3em}}
\date{}

\begin{document}

\maketitle

\section*{Abstract}

Decision-makers abhor uncertainty, and it is certainly true that the less there is of it the better. However, recognizing that uncertainty is part of the equation, particularly for deciding on environmental policy, is a prerequisite for making wise decisions. Even making no decision is a decision that has consequences, and using the presence of uncertainty as the reason for failing to act is a poor excuse. Statistical science is the science of uncertainty, and it should play a critical role in the decision-making process. This opinion piece focuses on the summit of the knowledge pyramid that starts from data and rises in steps from data to information, from information to knowledge, and finally from knowledge to decisions. Enormous advances have been made in the last 100 years ascending the pyramid, with deviations that have followed different routes. There has generally been a healthy supply of uncertainty quantification along the way but, in a rush to the top, where the decisions are made, uncertainty is often left behind. In my opinion, statistical science needs to be much more pro-active in evolving classical decision theory into a relevant and practical area of decision applications. This article follows several threads, building on the decision-theoretic foundations of loss functions and Bayesian uncertainty.

\section{Introduction}

Making a decision under almost any circumstance comes with uncertainty. For example, a country's investment in carbon capture and storage (CCS) comes with only a degree of certainty that the technology can be applied \textit{at scale}. Australia's current commitment to ``net zero emissions by 2050'' assumes that CCS, along with other technologies not even conceived of yet, will contribute substantially to reducing Australia's greenhouse gas emissions. In the Australian Government's 2021 plan, there is a prediction that ``... future technology breakthroughs will reduce emissions by ... 15 per cent by 2050'' \citep{Commonwealth2021}. The remaining 85\% appears to be less uncertain in the Australian Government's technology-driven plan: 40\% comes from driving down technology costs and accelerating their deployment at scale \citep{TechnologyInvestmentRoadmap2021}; 20\% comes from emissions reductions already made since 2005; 15\% comes from global technology trends; and 10\% comes from high-integrity offsets. A decision was made by the Australian Government in 2021 to rule out taxes and legislative mechanisms as part of achieving net zero carbon emissions by 2050, in line with the ruling Liberal-National Party (LNP) coalition's conservative agenda. (In the federal election of May 2022, the LNP were defeated by the Australian Labor Party, who promised to accelerate the passage to net zero carbon emissions -- new decisions by the new Australian Government are forthcoming.) 

Policy-makers make decisions based on data, modeling (depending on assumptions), predictions (from the modeling), and expert opinions (sometimes colored by the politics of the day). Where is there \textit{not} uncertainty in this process, yet how often is it made transparent? Even the emissions reductions of 20\% claimed between 2005 and 2021 in the Australian Government's plan is based on industry and state-government self-reporting; deeper in the plan, a figure to one decimal place of 20.8\% is given, indicating high certainty but based on figures whose uncertainty has almost certainly (!) not been quantified. 

Statistical science is the science of uncertainty, which includes its quantification, its estimation, and its consequences for inference in scientific models and the decisions that follow. Statistical scientists have almost always focused on how uncertainty and variability affect inference on key parameters and (summaries of) scientific processes; however, much less of their literature is devoted to the effect of uncertainty on decisions. Sometimes decision-makers see uncertainty as a reason for lack of action, but that too is a decision, and it has consequences. Sometimes the uncertainty of inference on an important environmental parameter is provided but ignored by the decision-maker, and only the estimate remains, to be seen as precise as the number of digits quoted. 

Science (environmental, medical, social, statistical, etc.) in its service to society should be better integrated into the peak of the \textit{knowledge pyramid}, where decisions are made: At the base of the pyramid are data; at the next level up is information obtained by exploring the data for structure; the information is then converted into knowledge at the next level by modeling the uncertainty/variability and inferring etiology; and finally, at the peak, decisions are made. Uncertainty quantification should be involved at all levels, and the consequences of ignoring it in environmental problems can be serious. For example, deciding to evacuate neighbourhoods and townships faced with possible flooding is based on meteorological and hydrological forecasts. Uncertainty of both forecasts should jointly inform evacuation decisions by government agencies; communities cannot be kept safe ``on average."

In the rest of this article, I shall discuss decision-making in the presence of uncertainty, opining that its future is in addressing the complex decisions that arise in uncertain environments. Section 2 reviews several approaches that have been taken in this area. Section 3 makes the case that decision theory and its applications have much to offer decision-makers, with its quantification of uncertainty through a posterior probability distribution and its quantification of the consequences of possible actions through a loss function. Several classes of loss functions are presented as alternatives to squared-error loss, including asymmetric loss functions that are key for making environmental decisions. Section 4 gives an example of how one such loss function can be calibrated. Finally, Section 5 gives concluding remarks and discusses new research directions.

\section{A selection of criteria to help decision-makers}

Before committing effort and resources to making a decision, ask first ``What is the question that the decision will answer?'' In the example given in Section 1, the question is, ``How will Australia achieve net zero carbon emissions by 2050?'' It took about five years from the time of the COP21 Paris Agreement for the Australian Government to seriously address that question. The best questions focus the decision-maker to provide the ``best'' answers/decisions. Providing several ``good'' but sub-optimal answers is not all that bad, but lack of transparency through a ``good-better-best'' criterion is.

In that spirit, I am arguing for clarity of assumptions, a statement of the criteria and a common num\'{e}raire against which all competing decisions are evaluated, an error budget that sets out sources of uncertainty, a statistical model that quantifies the uncertainties, and inferences that incorporate uncertainty quantification. Not just any model will do and, guided by the error budget, decision-making in the presence of uncertainty is in my opinion best served with a well fitted Bayesian hierarchical model (BHM). Complexity in underlying processes can often be captured by a computer model \cite[e.g.,][]{Santor2010} or what has been referred to as a ``digital twin.'' When building a BHM, the digital twin or some approximation to it could be used as the first moment of the hidden process. 

Complexity is often the case in the environmental sciences, where there are many questions asked by many stakeholders. Taking a BHM approach means that all the answers can be found somewhere in a (high-dimensional) posterior distribution. This article addresses, \textit{inter alia}, how one might organize the knowledge contained in the posterior distribution in order to make an optimal decision.

Section 3 presents a decision-theoretic approach to inference, with an emphasis on \textit{prediction} of unknown random quantities, rather than on estimation of unknown parameters. A relatively new approach to making the difficult decisions one sees in environmental applications is \textit{VOI} (\textit{Value of Information}), which aims to understand what is gained by taking additional types and amounts of data \cite[e.g.,][]{Eidsvik2015}. Suppose ``value'' $V$ is measured by a particular forecast's accuracy (economic, meteorological, epidemiological, etc.), but there is a suggestion that an extra data set might be incorporated to improve the accuracy. Then, 
$$
VOI \equiv E\!\left(V \mid \text{existing and extra data}\right) - E(V\mid \text{existing data}).
$$
Of course, a threshold will need to be specified to decide whether the additional data adds enough value. In some instances, the extra data set is costly and requires a cost-of-sampling function to offset the data's added value. As discussed in Section 5, incorporating the cost of data is an important part of sampling design; not all sensors are cheap, and ``found'' data can be of variable quality and value.

When the measurement error of the data is assumed to be zero, the information is deemed ``perfect'' and \textit{VOPI} (Value of Perfect Information) is an aspirational \textit{VOI} value: Then a proposed data set's \textit{VOI} can be compared to the corresponding \textit{VOPI} \cite[e.g.,][]{Lark2022}. Use of \textit{VOI} to make decisions is a relatively new approach, and practical issues are still being investigated. 

Consider the basic problem of deciding between competing models, $M_1, ..., M_m$. One classical approach is to use \textit{Bayes factors} \cite[e.g.,][]{Berger1985}. Define $\pi_k \equiv \mathrm{Pr}(M_k)$, the prior probability that $M_k$ is the correct model, and assume that one of the $k=1,...,m$ models is correct. Informed by data $z$, the posterior probability that $M_k$ is the correct model is 
$$
p(k \mid z) \equiv \mathrm{Pr}(M_k \mid z) \propto \mathrm{Pr}(z \mid M_k)\pi_k; k = 1, ..., m.
$$
Then $M_k$ could be compared to any of the other models, here $M_j$, by considering the Bayes factor,
$$
\frac{\mathrm{Pr}(z \mid M_k)}{\mathrm{Pr}(z \mid M_j)} = \frac{p(k \mid z)}{p(j\mid z)} \times \frac{\pi_j}{\pi_k},
$$
and one could choose the optimal model, $\delta_{BAF}^*(z) \equiv \mathrm{argmax} \left\{ p(k\mid z)/\pi_k: k = 1, ..., m\right\}$.

However, choosing the wrong model has differential consequences, depending on what is the correct model and which model was chosen. Another classical criterion is the \textit{expected posterior loss}. Suppose those consequences are quantified by a decision table, which is an $m\times m$ matrix of losses $(L_{jk})$ where, for $j,k = 1, ..., m$, $L_{jk}$ is the loss incurred by choosing model $M_j$ when $M_k$ is the correct model, and $L_{jj} = 0$. A decision to choose model $M_j$ has expected posterior loss, 
\begin{equation}
E_{PL}(j \mid z)\equiv \sum_{k=1}^m L_{jk} p(k\mid z); j = 1,...,m.\label{eqn:2.1}
\end{equation}
Then, minimizing the $E_{PL}$ criterion \eqref{eqn:2.1} over $j \in \{1, ..., m\}$ results in an optimal decision, which is to choose the model, $\delta_{EPL}^*(z) \equiv \mathrm{argmin} \left\{E_{PL}(j \mid z): j = 1,...,m\right\}$. 

These two decisions, $\delta_{BAF}^*$ and $\delta_{EPL}^*$, are only indirectly related, as can be seen if it is assumed that $L_{jk}$ is the indicator function, $I(j \neq k)$, sometimes known as the $0$-$1$ loss function. Then $E_{PL}(j\mid z) = 1 - p(j\mid z); j = 1,..., m$, and hence $\delta_{EPL}^*(z) = \mathrm{argmax} \left\{p(j \mid z): j = 1, ..., m\right\}$. Therefore, under a uniform prior (i.e., $\pi_j = 1/m$ for $j=1,...,m$), $\delta_{BAF}^* \equiv \delta_{EPL}^*$; otherwise, the two optimal decisions are different.
This most basic of decision problems, to decide between models $M_1, ..., M_m$, was chosen to illustrate that different criteria may lead to different optimal decisions. Since the method of Bayes Factors optimizes the (frequentist) likelihood, my preference is for a method directly based on the (Bayesian) posterior distribution, such as the expected posterior loss given by \eqref{eqn:2.1}. This is reinforced by the conclusions of \cite{Kass1995}, that Bayes factors should be avoided when the prior is not uniform.

Decisions come with consequences (losses), and it is important to capture loss along with the uncertainty. The essential property of a loss function is that it be bounded from below; hence, without loss of generality, it can be assumed to be non-negative. If one is to use expected posterior loss as a criterion, then it must also be assumed that its expectation with respect to the posterior probability measure exists (an assumption that involves the behavior of both loss and uncertainty). Uncertainties expressed through a BHM are traditionally made up of parametric conditional probabilities. A different emphasis is given in the book by \cite{Smith2010}, who features the Bayes networks that support a BHM but devotes less to the quantification of its conditional probabilities.

We shall see in the next section that the expected posterior loss is a way to account for the high-impact, low-probability scenarios seen in  the presence of extreme environmental threats. In my opinion, while statistical science has devoted much of its effort to the construction and properties of \textit{distribution functions for inference}, it has spent far less effort on developing \textit{loss functions for decisions}. Both are needed!

\section{From decision theory to decision applications}

Uncertainty quantification based on conditional-probability models offers a framework for optimal decision-making. In particular, hierarchical statistical modeling effectively separates out the data from the latent process and the parameters. In this article, I focus principally on decisions about (functions of) random process(es). Decision theory provides a way to determine an optimal predictor $\delta^*(z)$ of the random quantity $Y$ based on data $z$ in the presence of uncertainty in both $Y$ and $z$. 

Very simplistically, consider a set of actions $\{a^{(1)}, a^{(2)}, ...\}$ and a loss function $L$, where $L$ quantifies the ``balance sheet'' of costs and benefits that depend on both the true quantity $Y$ and an action $a$ that could be taken by the decision-maker; hence, write it as $L(a, Y)$. This allows action $a^{(1)}$ to be compared to action $a^{(2)}$ for each possible realization $y$ of $Y$. The action with the lower loss is preferred, but a uniform ordering for all $y$ may not be achievable. In what follows, a solution based on decision theory is presented. For the case where a multivariable decision is needed, Section 3.2 proposes an eigenanalysis that reduces the problem to making independent decisions in orthogonal eigenspaces. Sections 3.3 and 3.4 develop new loss functions of action $a$ and random variable $Y$ that give the decision-maker plenty of alternatives to the ubiquitous squared-error loss function, $L_{SEL}(a, Y) \equiv (a - Y)^2$.

\subsection{Prediction of a random quantity is a decision}

Consider the BHM defined by a sequence of conditional-probability models, as follows. First, the \textit{data model} is the conditional distribution, $f(z\mid y)$, of (potentially multivariate) data $z$ given the random process $Y = y$. Second, the \textit{process model} is the distribution $\pi(y)$, which quantifies the uncertainty in the latent process $Y$. When $Y$ is partially known through physical laws, the data model and the process model together define a physical-statistical model \cite[e.g.,][]{Kuhnert2014}. Now the distributions $f(z\mid y)$ and $\pi(y)$ are implicitly conditional on parameters that are notated here as $\theta$. Then third, the \textit{parameter model} is the prior probability distribution of the unknown parameters $\theta$.

Some realisations are probabilistically more likely than others given the data $z$, and here the presence of the posterior distribution of $Y$ is critical. (Note that the posterior $p(y\mid z)\propto f_{12}(y, z)$, where $f_{12}$ is the joint distribution, or uncertainty, of the unknowns $Y$ and $z$.) The classical approach taken in decision theory is to multiply the loss times the posterior density evaluated at a given $y$ and then to integrate over $y$ \cite[e.g.,][]{Berger1985}. That is, compute the \textit{expected posterior loss},
\begin{equation}
E_{PL}(a\mid z) \equiv \int L(a, y)p(y|z) \mathrm{d}y = E(L(a, Y)\mid z), \label{eqn:3.2}
\end{equation}
which is a function of $z$ with no dependence on $y$. Then \eqref{eqn:3.2} is a posterior weighted average of $L(a, y)$ over $y$, and the predictor $a^{(1)}$ is preferred to the predictor $a^{(2)}$ if $E_{PL}(a^{(1)}\mid z)\leq E_{PL}(a^{(2)}\mid z)$. Hence, under criterion \eqref{eqn:3.2}, a uniform ordering of $L(a^{(1)}, y)$ versus $L(a^{(2)}, y)$ for all $y$, is not required.

Note that once $L$ is specified, the set of all possible losses, $\{L(a, y)\}$ does not change. However, the posterior distribution $p(y\mid z)$ changes according to the data $z$ observed and the modeling assumptions made about  the uncertainty of $Y$ and $z$. It is used in \eqref{eqn:3.2} to weight the possible realisations of $L(a,y)$ for a fixed $a$. Then $a$ is varied until a decision $\delta^*(z)$ is found that minimizes \eqref{eqn:3.2}. That is, $\delta^*(z)$ is the best decision about $Y$ based on the optimization criterion $E_{PL}$ given by \eqref{eqn:3.2}. Notice that $Y$ is quite general and only needs to be defined on a measurable space. For example, the case of multivariate spatio-temporal prediction is discussed in Section 3.2, and the case of prediction in regression is presented in \cite{Kowal2021}.

Now suppose a different question is asked that involves making inference not on $Y$ but on a given functional $g(Y)$. A different loss function might be used but, for simplicity, suppose here that the same one is used. Then the integrand in \eqref{eqn:3.2} is replaced with 
\begin{equation} 
\int {L(a_{g}, g(y)) p(y\mid z)} \mathrm{d}y,  \label{eqn:3.3} 
\end{equation}     
where $a_{g}$ is a generic action to take about $g(Y)$. Minimizing \eqref{eqn:3.3} with respect to $a_g$ yields an optimal decision, $\delta^{*}_{g}(z)$, about $g(Y)$. It is tempting but not optimal to predict $g(Y)$ with $g(\delta^*(z))$, since that predictor does \textit{not} minimize \eqref{eqn:3.3} unless $g(y)$ is linear in $y$. For example, if $g(Y) = Y^2$ or $\exp\{Y\}$ or $I(Y > 0)$, then \eqref{eqn:3.3} needs to be minimized separately for each individual $g(\cdot)$. 

An obvious special case is where the random quantity $Y$ is a random variable; Sections 3.3, 3.4, and 4 consider this in detail. In Section 3.2 to follow, $Y$ is a multivariate spatio-temporal process, a situation that often arises when making decisions about the environment. 

\subsection{Prediction of many variables}

Applications of decision theory need to go beyond choosing between a finite number of models (Section 2) and prediction of a random variable (Sections 3.3, 3.4, and 4). All things are more-or-less related in some grand way, but science investigates which variables are most related and how. Questions in environmental science are often answered by first asking ``Where?'' and ``When?''. Consequently, prediction of multivariate spatio-temporal processes in the presence of uncertainty is an important problem that requires not only careful statistical modeling and efficient computations, but it also requires loss functions that emphasize the subspaces where critical action is needed. 

Let $Y\equiv \{\mathbf{Y}(\mathbf{s}; t): \mathbf{s} \in D_s, t \in D_t\}$ denote a vector-valued spatio-temporal process defined on the spatio-temporal domain $D_s \times D_t$, where $\mathbf{Y}(\mathbf{s}; t) \equiv (Y_i(\mathbf{s}; t): i = 1, ..., N)'$, and $N$ is a positive integer. It is important to recognise that environmental processes may be multivariate and that the units of various components may be very different. Let $\mathbf{R}$ denote the $N\times N$ \textit{correlation matrix} of $\mathbf{Y}(\mathbf{s}; t)$, where it is assumed that this matrix is positive-definite and invariant over $D_s \times D_t$ (i.e., it does not depend on $(\mathbf{s}; t)$). However, non-stationary spatio-temporal dependence in $Y$ is allowed, since the invariance assumption only applies to the correlations and not the means and variances. In practice, $\mathbf{R}$ could be estimated from noisy observations on $Y$. 

Consider the spectral decomposition, $\mathbf{R} = \mathbf{P} \bm{\Lambda}\mathbf{P}' = \sum_{i=1}^N \lambda_i \mathbf{P}_i \mathbf{P}_i'$,  where $\bm{\Lambda} \equiv \text{diag}(\lambda_1, ..., \lambda_N)$ is a $N \times N$ diagonal matrix of decreasing eigenvalues, and the $N \times N$ matrix $\mathbf{P} \equiv (\mathbf{P}_1, ..., \mathbf{P}_N)$ is made up of the corresponding eigenvectors (satisfying $\mathbf{P}'\mathbf{P} = \mathbf{I})$. To build a loss function, one could project $\mathbf{Y}(\mathbf{s}; t)$ and a generic predictor $\mathbf{a}(\mathbf{s}; t)$ onto each of the eigenspaces. That is, for $i = 1, ..., N$, define $x_i(\mathbf{s}; t) \equiv \mathbf{P}_i'Y(\mathbf{s}; t)$ and $b_i(\mathbf{s}; t) \equiv \mathbf{P}_i' \mathbf{a}(\mathbf{s}; t)$. Then the loss function $L_i(b_i(\mathbf{s}; t), x_i(\mathbf{s}; t))$, acts on the $i$-th orthogonal eigenspace, and the expected posterior loss $L_i$ can be minimized with respect to $b_i(\mathbf{s}; t)$ to yield $\gamma_i^*(\mathbf{s}; t)$. 

It is easy to see that the predictor of $\mathbf{Y}(\mathbf{s}; t)$,
\begin{equation}
\bm{\delta}^*(\mathbf{s}; t) \equiv \mathbf{P}(\gamma_1^*(\mathbf{s}; t), ..., \gamma_N^*(\mathbf{s}; t))', \label{eqn:3.4}
\end{equation}
minimizes, with respect to $\mathbf{a}(\mathbf{s}; t)$, the expected posterior loss given by 
\begin{equation}
E_{PL}(\mathbf{a}(\mathbf{s}; t)\mid z) \equiv E\!\left(\sum_{i=1}^N L_i(\mathbf{P}_i'\mathbf{a}(\mathbf{s}; t), \mathbf{P}_i'\mathbf{Y}(\mathbf{s}; t))\mid z\right). \label{eqn:3.5}
\end{equation}
Obvious examples of $L_i$ are the squared-error loss function, $L_i(b_i, x_i) = w_i(b_i - x_i)^2$, and the $0$-$1$ loss function, $L_i(b_i, x_i)\equiv w_iI(b_i - x_i \neq 0)$, where $w_i > 0$. Since $\mathbf{P}_1, ..., \mathbf{P}_N$ are ordered according to the eigenvalues $\lambda_1 \geq \lambda_2\geq ...\geq \lambda_N > 0$, the coefficients $\{w_i\}$ in these examples could be given by a monotonic increasing function of $\{\lambda_i\}$. To my knowledge, the spatio-temporal predictor \eqref{eqn:3.4}, obtained by minimizing \eqref{eqn:3.5}, is new and has not been investigated.

\subsection{Building loss functions}

Classical presentations of decision theory \cite[e.g.,][]{Berger1985} consider the loss function as given and then use the expected loss under the joint distribution of $Y$ and $z$ (expected joint loss), as the optimization criterion. It is straightforward to show that the consequent expected posterior loss (i.e., conditional on $z$) yields the same optimal predictor when minimized. However, the expected posterior loss as a function of action $a$ may order differently when $z = z^{(1)}$ than when $z = z^{(2)}$. This is the nature of the uncertainty, and it can be accounted for coherently through the joint distribution of $Y$ and $z$, $f_{1,2}(y, z) = p(y\mid z) f_2(z)$, where $f_2(z)$ is the marginal distribution of $z$, and recall that $p(y\mid z)$ is the posterior distribution of $Y$.

The mathematical properties of a loss function are remarkably mild: For $a$ belonging to the space of actions $\mathcal{A}$ and $y$ belonging to the space of predictands, $\mathcal{Y}$, the loss function $L(a, y)$ has domain given by $(a, y) \in \mathcal{A} \times \mathcal{Y}$ and range given by $\mathbb{R}^{+} \cup \{0\}$, the non-negative real line. Furthermore, $\mathcal{Y} \subset \mathcal{A}$ and when $a = y$, $L(y, y) = 0$. Finally, the expectation in $E_{PL}$ has to be well defined in order for it to be used as an optimization criterion.

Often, the space of possible actions $\mathcal{A}$ is a metric space with metric $d(\cdot, \cdot)$, and $L$ is given by $L_{MTC}(a, y) = m(d(a, y))$, where $m(\cdot)$ is a monotonic increasing function on the non-negative real line. For example, when $Y$ is a random variable, \textit{squared-error loss} is given by $L_{MTC}^{(2)}(a, y) = (|a - y|)^2$, where $d(a, y) = |a - y|$ and $m(x) = x^2$.  This generalizes to the loss function,
\begin{equation}
L_{MTC}^{(\rho)}(a, y) = (|a - y|)^\rho; \rho \in (0, \infty).\label{eqn:3.6}
\end{equation} 
Then absolute-deviation loss is $L_{MTC}^{(1)}$, and the $0$-$1$ loss function is obtained in the limit as $\rho \rightarrow 0$ from above: $L_{MTC}^{(0)}(a, y) \equiv I(a - y \neq 0)$.

It is straightforward to show that by substituting $L_{MTC}^{(2)}$, $L_{MTC}^{(1)}$, and $L_{MTC}^{(0)}$ into \eqref{eqn:3.2} and minimizing it with respect to $a \in \mathbb{R}$, one obtains the optimal predictor of $Y$ as, respectively, the posterior mean, the posterior median, and the posterior mode. These three predictors actually coincide for relatively small families of symmetric unimodal posterior densities (e.g., scale mixtures of Gaussian densities). For unimodal but skewed densities, there is a well known ``mode, median, and mean'' inequality \cite[e.g.,][]{Groeneveld1977}, which illustrates that choice of loss function can move the optimal predictor of $Y$ substantially in one direction or another.

In the rest of Section 3, $Y$ is a \textit{random variable} to be predicted, for which classical, less well known, and new loss functions are discussed and compared. Asymmetry in a loss function, where $L(a, y)$ is different for $a < y$ than for $ a > y$, captures the reality that the loss of an underprediction can be greater than the loss of an overprediction (or \textit{vice versa}), such as for prediction of a river's crest-height at a time of heavy rainfall in its catchment. An example of an asymmetric loss function is:
\begin{equation}
L_{QTL}^{(q)}(a, Y) = (a - Y)(I(a - Y > 0) - q); q \in (0, 1),\label{eqn:3.7}
\end{equation}
where $I(\cdot)$ is the indicator function. Upon minimizing the $E_{PL}$ given by $E(L_{QTL}^{(q)}(a, Y)\mid z)$, with respect to $a$, it is easily seen that the optimal predictor is the $q$-th posterior quantile. Further, $L_{QTL}^{(0.5)}(a, Y)  = L_{MTC}^{(1)}(a, Y)$, and $q = 0.5$ results in the only symmetric loss function in the class \eqref{eqn:3.7} generated by $q \in (0, 1)$. Choosing $0.5 < q <  1$ yields larger loss when $a < Y$ (i.e., underprediction) than when $a > Y$, which is appropriate when deciding how best to react to a flooding threat. For example, suppose a decision has to be made whether to evacuate low-lying neighbourhoods (in the short-term) or deciding how high to build a town's levy (in the long-term). This amounts to predicting the indicator function, $I(Y > \kappa)$, for different thresholds $\kappa > 0$. In the spatial context, such indicator functions define exceedance sets \citep{Cressie2020}.

Another asymmetric family is defined by the LINEX loss function (\citealp{Varian1975}; \citealp{Zellner1986}; \citealp[p. 108]{Cressie1993}),
\begin{equation}
L_{LNX}^{(\psi)}(a, Y) \equiv \exp\{\psi (a - Y)\} - \psi(a - Y) - 1; \psi \in (-\infty, \infty).\label{eqn:3.8}
\end{equation}
Choosing $\psi < 0$ gives larger loss when $a < Y$ (i.e., underprediction). Upon substituting $L^{(\psi)}_{LNX}(a, Y)$ into \eqref{eqn:3.2} and minimizing it with respect to  $a\in \mathbb{R}$, one obtains the optimal predictor of $Y$:
$$
\delta_{LNX}^*(z) = (-1/\psi)\log(E(\exp\{-\psi Y\}\mid z)); \psi \in (-\infty, \infty)
$$
which, by Jensen's inequality, is larger than $E(Y\mid z) \text{ for } \psi < 0$. That is, when $\psi < 0$ in the LINEX loss function \eqref{eqn:3.8}, the optimal predictor is larger than the classical posterior-mean predictor. While the two asymmetric classes, $\{L_{QTL}^{(q)}: q \in (0, 1)\}$ given by \eqref{eqn:3.7} and $\{L_{LNX}^{(\psi)}: \psi \in (-\infty, \infty)\}$ given by \eqref{eqn:3.8}, are quite different in terms of their differentiability at $a = Y$ and their shape as a function of $(a - Y)$, qualitatively they each have the ability to assign asymmetric losses that yield intuitively reasonable predictors.

There are other quite straightforward ways to define loss functions. Let $f(u; \omega)$, for parameters $\omega \in \Omega$, denote a parametric bounded density function of $u$ such that $f(u; \omega) \leq f(0; \omega)$, for $u \in \mathbb{R}$. Then the so-called potential function can also be used to define a loss function,
\begin{equation}
L_{PTL}^{(\omega)}(a, Y) \equiv -\log(f(a - Y; \omega)) + \log(f(0;\omega)); \omega \in \Omega \label{eqn:3.9}
\end{equation}
which, depending on $f(\cdot; \omega)$, may or may not be symmetric as a function of $a$ around $Y$. Note that $L_{PTL}^{(\omega)} \geq 0$ and $L_{PTL}^{(\omega)}(Y, Y) = 0$. A symmetric class is defined by the generalized Gaussian density, $f(u; \omega) = (1/2)(\omega/\Gamma(\omega^{-1}))\exp\{-|u|^\omega\}; \omega \in (0, \infty)$, and so \eqref{eqn:3.9} becomes
\begin{equation}
L_{PTL}^{(\omega)}(a, Y) \equiv |a - Y|^{\omega}; \omega \in (0, \infty).\label{eqn:3.10}
\end{equation}
Clearly, the same family of loss functions, $\{L_{MTC}^{(\rho)}: \rho \in (0, \infty)\}$ given by \eqref{eqn:3.6} and $\{L_{PTL}^{(\omega)}: \omega \in (0, \infty)\}$ given by \eqref{eqn:3.10}, can be obtained by going down different paths of construction.

A straightforward way to generate an asymmetric loss function from a symmetric one, $L$ say, is to take positive powers, $(L)^p$ for $p \in (0, \infty)$, or to exponentiate, $\exp\{L\} - 1$, and there are many other ways given the mild requirements for a function of $(a, Y)$ to be a loss function. For example, the (weighted) sum of two loss functions is a loss function (such as \eqref{eqn:3.5} in Section 3.2), as is the product of two loss functions. Thus, taking all weighted sums of products of some well known loss functions generates a very broad class. 

While loss functions appear to be plentiful, once chosen, there may be a problem to minimize the criterion $E_{PL}$ given by \eqref{eqn:3.2}, for what could be quite an unusual loss function. For example, the loss function, $L(a, Y) \equiv L^{(2)}_{MTC}(a, Y) + L_{MTC}^{(0)}(a, Y)$, is zero (as it should be) but discontinuous at $a = Y$. It assigns substantial loss for $a$ infinitesimally close to $Y$ and even more substantial additive squared-error loss for $a$ far away from $Y$. Clearly, loss functions can be pieced together to avoid discontinuities and even defined to be $0$ in a measurable region of $\mathcal{A}$ that includes $a = Y$; again, minimizing $E_{PL}$ may prove problematic.

Having rich classes to draw from is essential, but calibrating loss functions to similar past environmental threats or to simulations of $Y$ and $z$ based on a digital twin has, to my knowledge, not been formalized. Following historic rain events in February 2022 in central Eastern Australia, towns along major rivers were inundated. Many houses in neighbourhoods were flooded up to two storeys, and thousands of residents were displaced. On 2 March 2022, Andrew Hall, CEO of the Insurance Council of Australia (http://insurancecouncil.com.au), was interviewed on an Australian Broadcasting Corporation news station. He stated that of all the money spent on these types of disasters, only 3\% is typically spent on prevention/planning. The remaining 97\% is spent on clean-up. Calibrating this to the quantile loss functions, $\{L_{QTL}^{(q)}: q \in (0, 1)\}$, one might predict a river's crest-height $Y$ with the $0.97$ posterior quantile of $p(y\mid z)$, to help decision-makers decide whether or not to evacuate low-lying neighborhoods. In Section 4, I show how this type of calibration can be used in the LINEX class of asymmetric loss functions given by \eqref{eqn:3.8}.

\subsection{Loss functions not based on displacement}

A generic loss function $L(a, Y)$ does not have to be a function of the displacement, $a - Y$, but often it is the default choice. All the examples given in Section 3.3 (i.e., $L_{MTC}$, $L_{QTL}$, $L_{LNX}$, $L_{PTL}$) are functions of displacement. One easy way to generalise any loss function is to define a weighted version, where the weight is a function of $Y$. If $L(a, Y)$ is a function only of displacement, its weighted version, $L_w(a, Y)$, will lose that property:
\begin{equation}
L_{w}(a, Y) \equiv w(Y)L(a, Y), \label{eqn:3.11}
\end{equation}
where $w(\cdot) > 0$. Then substituting $L_\omega$ into \eqref{eqn:3.2} yields $E_{PL}$ given by $\int w(y)L(a, y) f(y \mid z) \mathrm{d}y \propto \int L(a, y) p_w(y \mid z) \mathrm{d}y$, where $p_w(y\mid z) \propto p(y\mid z)w(y) \propto f(z \mid y) \pi(y) w(y)$. Analytically, the effect of using the weighted version \eqref{eqn:3.11} is to modify the distribution $\pi(\cdot)$ and replace it with $\pi_w(\cdot) \equiv \pi(\cdot)\omega(\cdot)$. Then the weighted posterior is 
$$
p_w(y\mid z) = \frac{f(z\mid y)\pi_w(y)}{\int f(z\mid x)\pi_w(x) \mathrm{d}x} = \frac{w(y)p(y\mid z)}{\int w(x) p(x \mid z) \mathrm{d}x}.
$$

If \eqref{eqn:3.2} can be minimized based on $p(y\mid z)$, then an analogous result follows immediately based on $p_w(y\mid z)$. For example, under squared-error loss, where $L_{SEL}(a, Y) = (a - Y)^2$, the optimal predictor of $Y$ is $\delta_{SEL}^*(z) = E(Y\mid z)$. Hence, if $L_{SEL, w}(a, Y) \equiv w(Y)(a - Y)^2$, then the optimal predictor of $y$ is $\delta^*_{SEL, w}(z) \equiv E_w(Y\mid z)$, where the expectation $E_w$ is taken with respect to $p_w(\cdot \mid z)$. In a more complicated multivariate problem of finding hotspots among the latent spatial process, $Y\equiv \{Y(\mathbf{s}_1), ..., Y(\mathbf{s}_n)\}$ of $n$ small-area disease rates at spatial locations $\{\mathbf{s}_1, ..., \mathbf{s}_n\}$, \cite{Wright2003} used a weighted ranks squared error loss (WRSEL) function, where the weights were functions of the ranks of $\{Y(\mathbf{s}_i)\}$, and the data were the observed disease counts $z \equiv \{z(\mathbf{s}_1), ..., z(\mathbf{s}_n)\}$.

In some settings, $Y$ is positive, such as when it represents a random rate. Then loss may be better represented as a function of the ratio, $a/Y$, rather than of the displacement, $a - Y$. An obvious example derives  from the absolute percentage error, $L_{APE}(a, Y) \equiv |1 - a/Y|$. Goodness-of-fit tests are often expressed in terms of divergence measures that are functions of ratios. For example, the power-divergence between two non-negative sequences $\{b_h: h = 1, ..., H\}$ and $\{c_h: h = 1, ..., H\}$ is defined by \cite{Read1988} as, $P \equiv \sum_{h=1}^H c_h \phi_\lambda(b_h/c_h)$, where $\lambda \in (-\infty, \infty)$ and $\phi_\lambda(r) \equiv (\lambda(\lambda + 1))^{-1}\{(r^{\lambda + 1} - r) + \lambda(1 - r)\}$, for $r \in (0, \infty)$. Then a class of loss functions for $Y > 0$ can be defined as:
\begin{equation}
L_{PWD}^{(\lambda)}(a, Y) \equiv Y\phi_\lambda(a/Y); -\infty < \lambda < \infty,\label{eqn:3.12}
\end{equation}
where the members, $\lambda = 0, -1$, are given by their respective limits. A related family has been used by \cite{Cressie2022} to generalize kriging \citep{Matheron1963} to spatial prediction of a non-negative spatial process. Note that the loss \eqref{eqn:3.12} as a function of $a$ around $Y$ is asymmetric; for example, when $\lambda < 0$, underprediction incurs more loss than overprediction. 

Other classes that are loss functions of the ratio $(a/Y)$ can be found by selecting appropriate probability density functions from the exponential family of distributions. Consider the probability density,
\begin{equation}
h^\alpha(x) \equiv \exp\{-\alpha x\} l(x)/k(\alpha); x \geq 0,\label{eqn:3.13}
\end{equation}
where $\alpha > 0$ is a parameter, the normalizing constant $k(\alpha)$ is a function of $\alpha$ that guarantees the density integrates to 1, and $l(x)$ can be thought of as a base measure that controls the shape of the density. For example, consider the gamma distribution, which is in the exponential family; for a given $\nu > 1$, it has base measure given by $l^{(\nu)}(x) = x^{\nu-  1}$ and normalizing constant given by $k^{(\nu)}(\alpha) = \Gamma(\nu)/\alpha^\nu$. That is, define the family of gamma probabilitiy densities, $h_{GAM}^{(\alpha, \nu)}(x) \equiv \exp\{-\alpha x\} l^{(\nu)}(x)/k^{(\nu)}(\alpha); x > 0$, for $\alpha > 0$ and $\nu > 1$. 

The mode of the gamma density is at $x^{(0)} = (\nu - 1)/\alpha$, and hence in a similar manner to the definition \eqref{eqn:3.9}, the following loss function can be defined:
$$
L_{GAM}^{(\alpha, \nu)}(a, Y) = \log\left\{h_{GAM}^{(\alpha, \nu)}\left(\frac{\nu - 1}{\alpha}\right)\right\} - \log\left\{h_{GAM}^{(\alpha, \nu)}\left(\frac{\nu - 1}{\alpha}\left(a/Y\right)\right)\right\}; \alpha > 0, \nu > 1,
$$
which is a function of $a/Y$. Note that $L_{GAM}^{(\alpha, \nu)} \geq 0$ and $L_{GAM}^{(\alpha, \nu)}(Y, Y) = 0$. Since
$$
\log\left\{h_{GAM}^{(\alpha, \nu)}\left(\frac{\nu - 1}{\alpha}\left(a/Y\right)\right)\right\} = -(\nu - 1)\left(a/Y\right) + (\nu - 1)\log\left(\frac{\nu - 1}{\alpha}\right) + (\nu - 1)\log\left(a/Y\right),
$$
its derivative with respect to $a$ is $(\nu - 1)\left(Y^{-1} - a^{-1}\right)$. Hence, when $L_{GAM}^{(\alpha, \nu)}(a, Y)$ is substituted into \eqref{eqn:3.2} and the resulting $E_{PL}$ is minimized with respect to $a$, the optimal predictor is, for all $(\phi, \nu) \in (0, \infty) \times (1, \infty)$, 
\begin{equation}
\delta^*_{GAM}(z) \equiv \left\{E(Y^{-1}\mid z)\right\}^{-1}.\label{eqn:3.14}
\end{equation}

It is straightforward to show that \eqref{eqn:3.14} is also the optimal predictor based on $L_{PWD}^{(1)}(a, Y)$ given by \eqref{eqn:3.12}. Further, the weighted loss function, $YL_{GAM}^{(\phi, \nu)}(a, Y)$ results in the optimal predictor, $E(Y \mid z)$, which is also the optimal predictor for $L_{SEL}(a, Y) = L_{MTC}^{(2)}(a, Y)$ given by \eqref{eqn:3.6} and for $L_{PWD}^{(-1)}(a, Y)$ given by \eqref{eqn:3.12}. Clearly, very different loss functions can yield identitical optimal predictors and, hence, what discriminates between decisions in the presence of uncertainty is not only the predictor but also the quantification of its uncertainty via the assumed loss function (e.g., the minimized $E_{PL}$). There is also a notion of inefficiency caused by using a ``working'' loss function instead of the ``true'' loss function. Inefficiency of a decision can be defined, for example, via the working/true expected posterior losses. Inefficiencies also arise when a ``working'' probability model for $Y$ and $z$ are used instead of the ``true'' probability model; see \cite{Cressie2021}.

\section{Calibrating within a class of loss functions}

In Section 3, it was shown that loss functions abound, to capture circumstances where losses are not symmetric (e.g., squared-error loss) functions of displacement. When making decisions in the face of extreme events, asymmetric loss can be used to represent the often smaller cost of mitigation versus a major cost of recovery. 

In this section, the class of LINEX loss functions $\{L_{LNX}^{(\psi)}(a, Y): \psi \in (-\infty, \infty)\}$, given by \eqref{eqn:3.8}, quantifies this with a scalar tuning parameter $\psi$, where $\psi < 0$ recognizes that larger losses will be incurred from underprediction of $Y$ (e.g., $Y$ is a river's crest-height after an extreme amount of rainfall has occurred in the river's catchment). For $|\psi|$ small, a Taylor-series expansion results in (approximate) squared-error loss: $L_{LNX}^{(\psi)} \approx (\psi^2/2)(a - Y)^2 \propto (a - Y)^2$. However, as $\psi \rightarrow -\infty$, the loss when $a < Y$ (underprediction) grows exponentially compared to the loss when $a > Y$ (overprediction). Overprediction of the crest-height of a major river might lead to mass-evacuation orders that later could be seen as an unnecessary nuisance, but the alternative could be a very costly loss of material possessions/animals/crops/life. The asymmetric LINEX loss function can be tuned to make more conservative flood-related decisions by making $\psi$ more negative, resulting in a biased but conservative prediction of the crest-height. In this section, a way to calibrate $\psi$ is proposed.

Decisions are made in an uncertain environment, where here the uncertainty is expressed probabilistically, specifically through the joint distribution of $Y$ and $z$. We saw in Section 3.3 that the optimal predictor of $Y$ obtained from minimizing \eqref{eqn:3.2} with the LINEX loss function given by \eqref{eqn:3.8} is, for $\psi \in (-\infty, \infty)$,
\begin{align}
\delta_{LNX}^*(z) &= (-1/\psi)\log\!\left(E(\exp\{-\psi Y\} \mid z)\right)\nonumber\\
&\simeq (-1/\psi)\log\!\left(\exp\{-\psi \mu_{Y\mid z}\} + (\psi^2/2)\sigma^2_{Y\mid z} \exp\{-\psi \mu_{Y \mid z}\}\right)\nonumber\\
&= \mu_{Y\mid z} + (-1/\psi) \log(1 + (\psi^2/2)\sigma^2_{Y\mid z}), \label{eqn:4.15}
\end{align}
where the approximation relies on small $\sigma^2_{y\mid z}$ and is obtained using the delta method in terms of the conditional moments, $\mu_{Y\mid z} \equiv E(Y \mid z)$ and $\sigma^2_{Y\mid z} \equiv \text{var}(Y \mid z)$. A further approximation up to $O(\sigma^2_{Y\mid z})$ is provided by the Taylor-series expansion, $\log(1 + x) = x  + O(x^2)$; then from \eqref{eqn:4.15}, 
\begin{align*}
\delta_{LNX}^*(z) &\simeq \mu_{Y\mid z} - (1/\psi)\log(1 + (\psi^2/2)\sigma^2_{Y\mid z}) \simeq \mu_{Y\mid z} + (-\psi \sigma_{Y\mid z}/2)\sigma_{Y\mid z}~.
\end{align*}

For $\psi < 0$, the approximation shows that $\delta_{LNX}^*(z)$ conservatively overpredicts in relation to the posterior mean, $\mu_{Y\mid z} = E(Y \mid z)$. (It was shown in Section 3.3 that overprediction occurs for $\psi < 0$, no matter how large $\sigma^2_{Y\mid z}$ is.) It is now shown that the approximation allows $\psi$ to be chosen according to a back-of-the-envelope calculation: Based on the $0.97$ Gaussian quantile, 1.88, which is motivated by the discussion at the end of Section 3.3, we solve for $\psi$ in $(-\psi \sigma_{Y \mid z})/2 = 1.88$, giving $\psi = -3.76/\sigma_{Y \mid z}$. Note that the calibration could be improved if more were known about $p( Y \mid z)$, ideally its moment generating function. Unlike the quantile loss function, the LINEX loss function has all derivatives, including $\partial L_{LNX}/\partial a$. In general, it can be advantangeous in higher-dimensional problems to use loss functions that allow calculation of derivatives to arrive at an optimal decision in terms of $E_{PL}$ given by \eqref{eqn:3.2}.

\section{Conclusions and future research directions}

The previous sections have followed several threads of decision theory and demonstrated the importance of quantifying a decision $a$ in terms of a loss $L(a, Y)$, where $Y$ is the unobserved random quantity of interest. Data $z$ that are dependent on $Y$ are available, and the uncertainty in $Y$ and $z$ is expressed probabilistically through the joint distribution $f_{12}(y, z)$. The two essential components for decision theory and its applications are $L(a, y)$ and $f_{12}(y,z)$. Concentrating on $E_{PL}$ requires working with the posterior distribution, which is calculated via Bayes' Theorem, 
$p(y \mid z) = f_{12}(y,z)/f_2(z) = f(z \mid y)\pi(y)/f_2(z)$, where recall that $f(z\mid y)$ is the ``data model'' or the ``likelihood'', $\pi(y)$ is the ``process model,'' and $f_2(z)$ is the marginal distribution of $z$ (also called ``the normalizing constant'').

There are a number of key conclusions to draw: 
\begin{itemize}[noitemsep, topsep=1pt]
\item An optimal decision depends fundamentally on the criterion being optimized.
\item A decision comes with consequences (here expressed as losses), which depend on both the hidden process $Y$ and on the decision $\delta(z)$ about $Y$, where $z$ represents the data. Loss functions are plentiful.
\item The hidden process $Y$ is uncertain, and that uncertainty can be expressed according to a probability model $\pi(\cdot)$ for $Y$.
\item After obtaining data $z$, the process model $\pi(\cdot)$ is updated to the posterior distribution $p(\cdot \mid z)$ according to Bayes' Theorem, and that posterior should be used in the decision-making process, along with a loss function $L$.
\item The same optimal decision may come from quite different optimality criteria (defined by quite different loss functions).
\item The optimized criterion should not be forgotten and needs to be evaluated along with the optimal decision.
\item The optimality criterion is usually the expected posterior loss, $E_{PL}$ (but it does not have to be).
\end{itemize}
\par\bigskip
Classical decision theory has concentrated on two types of ``expected loss,'' namely \textit{Expected Joint Loss} ($E_{JL}$), sometimes referred to as Bayes risk or unconditional risk: 
$$
E_{JL}(L, \delta, f_{12}(\cdot, \cdot)) \equiv \int L(\delta(z), y)f_{12}(y, z) \mathrm{d}y\mathrm{d}z;
$$
and \textit{Expected Posterior Loss} ($E_{PL}$), sometimes referred to as conditional risk:
$$
E_{PL}(L, \delta(z), p(\cdot \mid z)) \equiv \int L(\delta(z), y) p(y\mid z) \mathrm{d}y.
$$

In my opinion, the word ``risk'' should \textit{not} be used to describe $E_{JL}$ and $E_{PL}$; ``risk'' should be used in a way that many decision-makers and the general public use it, namely as a probability. What's in a name? Everything! Traditionalists may not like it, but if one adopts, say, the abbreviations ``$E_{PL}$'' to replace ``conditional risk'' and ``$E_{JL}$'' to replace ``unconditional risk,'' it would avoid the profound ambiguity that currently exists where, in common parlance, ``risk'' often means ``probability'', not ``expected loss.''

The decision-theoretic approach optimizes $E_{JL}$ with respect to decision $\delta$ or, equivalently, it opimizes $E_{PL}$ with respect to $\delta(z)$. The two ingredients of $E_{PL}$ are combined first as a product, $L(\delta(z), y) \times p(y\mid z)$, which is a way of weighting the consequence of a decision about an event (no matter how catastrophic) with the risk/probability of that event (no matter how small), given the data $z$. Recall that these products are integrated over all possible events to form $E_{PL}$, which is then minimized with respect to $\delta(z) \in \mathcal{A}$, for each $z$, resulting in the optimal predictor $\delta^*(\cdot)$. 

A possible \textit{future direction} is to find other ways of combining $\{L(\delta(z), y): y \in \mathcal{Y}\}$ and $\{p(y \mid z): y \in \mathcal{Y}\}$. One way that I \textit{do not} advocate is minimizing the maximum loss, resulting in the so-called minimax decision, $\delta_{MML}(z) \equiv \mathrm{argmin}_{\delta(z)} \{\max_{y\in \mathcal{Y}}{L(\delta(z), y)}\}$, 
since it is blind to the uncertainty of the environment $Y$ by assuming all values of $y \in \mathcal{Y}$ are equally likely. However, minimizing the maximum posterior loss, \\$\delta_{MMP}(z) \equiv \mathrm{argmin}_{\delta(z)} \{\max_{y \in \mathcal{Y}}{L(\delta(z), y) \times p(y \mid z)}\}$, does combine a value of $y$ with its posterior probability but, instead of averaging, it takes the maximum. 

Just as the minimax criterion considers only $L$ and ignores $p$, the reverse sometimes happens where decisions are made based only on the posterior distribution $p$. For example, a ``yes/no'' decision might be based on a threshold $\kappa$ where ``yes'' corresponds to $\mathrm{Pr}(Y > \kappa) \geq 0.5$.  Some reverse engineering from Section 3.3 can be used to demonstrate that the implied loss function for the yes/no decision is an unweighted $0$-$1$ loss function applied to the probability model, $\pi_\kappa \equiv \mathrm{Pr}(Y > \kappa)$ and $1 - \pi_\kappa = \mathrm{Pr}(Y \leq \kappa)$. So under $0$-$1$ loss, one decides ``yes'' if $p_\kappa(z) \equiv \mathrm{Pr}(Y > \kappa \mid z) \geq 1 - p_\kappa(z)$, which is equivalent to $\mathrm{Pr}(Y > \kappa \mid z) \geq 0.5$. However, ``yes'' might be more costly than ``no,'' which can be captured with a loss function given by a decision table (described in Section 2).

In Sections 3.3 and 3.4, it was shown that much more is possible to capture asymmetric loss than simply using posterior quantiles. For example, consider the weighted 0-1 and quantile loss functions:
\begin{align*}
L_{MTC, w}^{(0)}(a, Y)&\equiv w(Y)I(a- Y\neq0)\\
L_{QTL, w}^{(q)}(a, Y)&\equiv w(Y)(a-Y)(I(a - Y > 0) - q),
\end{align*}
which are particularly useful in the spatial context. Now the \textit{risk maps} $\{\delta_{QTL}^{(q)}(z; \mathbf{s}): \mathbf{s} \in D \subset \mathbb{R}^d\}$, for various posterior quantiles over a spatial domain of interest $D$ (contained in $d$-dimensional Euclidean space $\mathbb{R}^d$, say), allow identification of at-risk regions in $D$. However, they do not use information about catastrophic losses that would be incurred should a threshold $\kappa$ be exceeded. That could happen by letting the weight $w(Y)$ depend on $\kappa$. 

There are other ways to combine loss and posterior probabilities. For example, recall the tail posterior probability, $\mathrm{Pr}(Y > \kappa \mid z) \equiv p_\kappa(z)$ for $\kappa \in (-\infty, \infty)$, which is sometimes called the \textit{tail risk}. Then for a given $\delta(z)$, consider the curve obtained by plotting $L(\delta(z), \kappa)$ on the vertical axis (from $0$ to $\infty$) versus $p_\kappa(z)$ on the horizontal axis (from $0$ to $1$), where the curve is a locus traced out by varying the threshold $\kappa$ over $\mathcal{Y}$. Each $\delta(z)$ generates a curve, and the lower bound of these curves over $\delta(z) \in \mathcal{A}$ can be plotted and represents the smallest possible loss, conditional on $z$. This lower envelope gives a benchmark against which predictors, such as $\delta_{EPL}(z)$, $\delta_{MML}(z)$, $\delta_{MMP}(z)$, and any others could be judged. The plot would also have visual appeal to decision-makers, who could assess the costs of different actions as the tail risk varies. 

Another approach is to define the probability model of uncertainty directly in terms of loss functions; see \cite{Bissiri2016}. In this framework, which is a form of generalized Bayesian inference, the likelihood (i.e., the data model) is replaced with a loss function. While the framework was originally laid out for performing inference on an unknown parameter of interest, it has since been expanded to prediction of an unknown random variable or process of interest \cite[e.g.,][]{Loaiza2021}.This ``generalized Bayes'' approach of \cite{Bissiri2016} is quite different from the classical decision-theoretic approach, since it combines the notions of loss and posterior distribution.

The decision-maker may convene an expert panel which is asked to reach a consensus about $Y$. This can be dangerous if a single $Y = y$ is declared: A consensus can be a decision that everyone agrees to collectively, but few believe in individually! The panel's distribution of beliefs about $Y$ could be considered to be an elicitation of the distribution $\pi(\cdot)$. However, even this could be unsatisfactory since, without data $z$, optimizing $E_{PL}$ results in a decision, $\delta_{EPL}^* \equiv \mathrm{argmin}_\delta \int L(\delta, y)\pi(y) \mathrm{d}y$, which is a predictor of $Y$ that is not informed by data. Of course, the expert panel might have commissioned the collection of data, but sometimes it does not happen because sampling is costly or (related) data are readily available. 

The cost of data $z$ should be accounted for. Suppose $z$ is made up of $n$ individual observations; to emphasize this, it is written here as $z_n$; consequently, the posterior distribution is written as $p(y\mid z_n)$. Let $c(n)$ denote a cost function that quantifies the cost of sampling. Now, the expert panel is faced with recommending a decision about the environment $Y$, but it first has to design a sample $z_n$ at a cost of $c(n)$ dollars. How large should $n$ be? 

To answer this question, define a new loss function based on the original $L(a, Y)$:
$$
L_{CST}(n, a, Y) \equiv \tau L(a, Y) + c(n),
$$
where $\tau > 0$ is an ``exchange rate'' that converts units of loss into dollars. The data have not been collected yet so, in terms of $E_{JL}$, the optimal $n$ is,
$$
n^* = \mathop{\mathrm{argmin}}_{n\geq 0} \left\{\tau \int L(\delta, y) f_{12}(y, z_n) \mathrm{d}y \mathrm{d} z_n + c(n)\right\}.
$$
For example, \cite{Cressie1989} solve this sample-size optimization problem for deciding between two simple hypotheses; and in a spatial design for regularly spaced ice-core sampling on a transect across Antarctica, \cite{Cressie1998} determined $n$ by optimizing the constant spacing between successive samples. The initial loss $L$ was not in units of dollars but expressed in terms of the kriging variance in a tranche of Antarctica that went from coast-to-coast through the South Pole, and $c(n)$ was linear in $n$ after an initial set up cost of $c(0) = c_0 > 0$.

So-called citizen science is one way to avoid the cost of sampling, but the uncertainty in the data conditional on $Y = y$, may be difficult to define or may be prohibitively large. This difficulty could be mitigated  if the citizen-scientists have been given a protocol and adhere to it (e.g., the breeding bird survey; see \citealp{Robbins1986}).

Decision theory informs decision applications and many applications involve multiple criteria. A possible \textit{future direction} is to develop a methodology that tackles situations where there are multiple stakeholders, each with their own loss function(s). In the context of making management decisions, to mitigate pollutants entering Australia's Great Barrier Reef lagoon from the Burdekin River catchment in North Queensland, \cite{Kuhnert2018} have put uncertainty at the centre of their approach. A variety of loss functions give a variety of optimal predictions along with uncertainty expressed through expected posterior losses. The authors combine these with a range of static and dynamic visualisations that the decision-maker and stakeholders can use to explore a shortlist of decisions and their consequences. The aim is that this will lead to a stakeholder-consensus loss function, although more likely it will quantify where the differences lie. However, even this limited outcome is beneficial since it crystalizes the consequences of any proposed decision. One way to construct a consensus loss is to take a weighted sum of everyone's loss function. Then the decision-maker's task would be to choose the weights after consulting with the stakeholders. A variety of loss functions may also be accompanied by a variety of posterior distributions, since statistical models of uncertainty are not ordained but involve a certain amount of judgement. Settling for a good decision rather than the best decision could be approached through the notion of \textit{bounded rationality} \citep{Simon1957}. Alternativey, each stakeholder could be considered to be a group member, and the problem could be approached through making a \textit{Bayesian group decision} \cite[e.g.,][]{Keeney2011}.

\cite{Stenitz2012} has proposed a non-probabilistic way to make decisions where different stakeholders have different loss functions. He called it \textit{Geodesign}, in which the environmental study asks six questions: How should the study area be described? How does the study area operate? Is the current study area working well? How might the study area be altered? What differences might the changes cause? How should the study area be changed? If uncertainty is incorporated into this series of questions, then it is clear that, currently, decision theory is not addressing them all. A possible \textit{future direction} is to expand decision theory so that it does. 

Now return to the basic problem discussed in Section 2, of choosing between $m$ models $M_1, ..., M_m$ with prior distribution $\{\pi_1, ..., \pi_m\}$. Based on data $z$, the posterior distribution is $p(k \mid z) = \mathrm{Pr}(M_k \mid z) \propto \mathrm{Pr}(z \mid M_k) \pi_k$, for $k = 1, ..., m$. Section 2 proposes minimizing $E_{PL}$ given by \eqref{eqn:2.1}, but it does not address the prediction problem. In what follows, an approach to optimal prediction, called \textit{Bayesian Model Averaging (BMA)}, is discussed \cite[e.g.,][]{Draper1995}. A possible \textit{future direction} is to fully integrate BMA into a decision-theoretic setting, and a start is made in what follows. Let $\hat{Y}_k$ be the optimal predictor of $Y_k$ under model $M_k$ where, for the moment, squared-error loss is assumed; hence, $\hat{Y}_k = E_k(Y_k \mid z)$, where $E_k$ denotes expectation under model $M_k$. The BMA predictor is now derived as an optimal decision that minimizes $E_{PL}$ for the squared-error loss function given by
\begin{equation}
L_{SEL}(a, Y_K) = (a - Y_K)^2,\label{eqn:5.16} 
\end{equation}
where $K$ is the unknown random index of the correct model, and $Y_K$ is the unknown predictand under the correct model. Then the criterion $E_{PL}$ given by \eqref{eqn:3.2} is:
$$
E((a - Y_K)^2\mid z) = \sum_{k=1}^m E_k((a - Y_k)^2 \mid z)p(k \mid z).
$$
Now differentiate the right-hand side with respect to $a$ and set the result equal to $0$. This yields, \\$\sum_{k=1}^m p(k \mid z) E_k(a - Y_k \mid z) = 0$ and, hence, the optimal predictor of $Y_K$ is,
\begin{equation}
\delta_{BMA}^*(z) \equiv \sum_{k=1}^m p(k \mid z)E_k(Y_k \mid z),\label{eqn:5.17}
\end{equation}
which is the BMA predictor. Squared-error loss may not be appropriate for all models. To allow for a diversity of loss functions, $L_1(a, Y_1), ..., L_m(a, Y_m)$, define the loss for prediction of $Y_K$ to be $L_K(a, Y_K)$, where $L_K$ is a random loss function. This generalises \eqref{eqn:5.16}, and hence $E_{PL}$ is simply, $\sum_{k=1}^m E(L_k(a, Y_k)\mid z)p(k \mid z)$. Consequently, the BMA predictor \eqref{eqn:5.17} generalizes to a posterior-weighted combination of possibly quite different individual predictors.

What should one do if none of the models is correct? The models are capturing the ``known unknowns,'' but the correct model may be one of the ``unknown unknowns.'' One way to fill this gap is to include an extra, maximum-entropy model \cite[e.g.,][]{Cressie2021}. Indeed, \citet[Ch. 11]{Le2006} proposed using only a maximum-entropy model for design and analysis because, in the design stage, the totality of questions that will be asked of the data is never known. 

Other possible \textit{future directions} include: optimal prediction of spatial processes $\{Y(\mathbf{s}): \mathbf{s} \in D\}$, including joint prediction of $Y$ at several spatial locations and block prediction of $Y(B) \equiv |B|^{-1} \int_B Y(\mathbf{s}) \mathrm{d}\mathbf{s}$; optimizing $E_{PL}$ or any other decision-theoretic criterion using a combination of simulation, emulation, and numerical methods; allowing artificial intelligence (AI) to make decisions in ways that may be difficult to formalize in terms of a loss function; and making decisions based on visualization and human intelligence (HI), particularly in a spatial setting using GIS (Geographical Information Systems) mapping capabilities. 

I conclude this article by opining that while diagnostics for fitting an appropriate statistical model of the uncertainties are commonly used, very little research has been done on diagnosing and validating the choice of a loss function. At the very least, \textit{post hoc evaluation} of the decisions made should be carried out as a matter of course. In that vein, a possible \textit{future direction} is to look at the consequences of using an inappropriate loss function whose $E_{PL}$ and $E_{JL}$ are not fit for purpose. That means the decision $\delta^*(z)$ that minimizes them may be non-optimal in terms of the ``true'' $E_{PL}$ that should have been used. While such optimized predictions may not be best, they may nonetheless be ``quite good,'' which can be determined by carefully designed sensitivity experiments evaluated in terms of the true $E_{PL}$ or the true $E_{JL}$.

\section*{Acknowledgments}

My thanks go to the editors of this special issue for the opportunity to be opinionated! I am also grateful to Alan Pearse for his dedicated help in preparation of this opinion piece. The research was supported by ARC Discovery Projects, DP190100180 and DP150104576.

\bibliographystyle{Chicago}
{\small \bibliography{NC_Envx_Opinion_WPS_09Sep22}}

\end{document}